\documentstyle[preprint,aps,epsf,floats,eqsecnum]{revtex}

\let\mdef\def
\mdef\etal{{\it et al.}}
\mdef\Tr{\mathop{\rm Tr}\nolimits}
\mdef\tr{\mathop{\rm tr}\nolimits}
\mdef\str{\mathop{\rm str}\nolimits}
\mdef\sdet{\mathop{\rm sdet}\nolimits}
\mdef\det{\mathop{\rm det}\nolimits}
\mdef\Str{\mathop{\rm Str}\nolimits}
\mdef\Sdet{\mathop{\rm Sdet}\nolimits}
\mdef\Det{\mathop{\rm Det}\nolimits}
\mdef\vslash{\mathop{\not \! v}}
\mdef\scr#1{{\cal #1}}
\mdef\eqb{begin{equation}}
\mdef\eqe{end{equation}}
\mdef\ecm{e\,{\rm cm}}
\mdef\MeV{{\rm \,MeV}}
\mdef\GeV{{\rm \,GeV}}
\mdef\mev{{\rm \,MeV}}
\mdef\gev{{\rm \,GeV}}
\mdef\qt{{\tilde q}}
\mdef\qbar{{\overline q}}
\mdef\qtbar{{\overline\qt}}
\mdef\phit{{\tilde\phi}}
\mdef\Sbar{{\overline S}}
\mdef\Tbar{{\overline T}}
\mdef\etat{{\tilde \eta}}
\mdef\mii{m_{ii}}
\mdef\mjj{m_{jj}}
\mdef\mij{m_{ij}}
\mdef\muu{m_{uu}}
\mdef\mss{m_{ss}}
\mdef\mdd{m_{dd}}

\tighten

\begin{document}

\preprint{\vbox{\hbox{JHU--TIPAC--97008}\hbox{hep-ph/9704426}}}

\title{Quenched Chiral Perturbation Theory for Heavy Baryons}
\author{George Chiladze}
\address{Department of Physics and Astronomy,
The Johns Hopkins University\\
3400 North Charles Street, Baltimore, Maryland 21218 U.S.A.}

\date{April 1997}
\maketitle

\begin{abstract} 

Heavy baryon chiral perturbation theory is extended to include the effects of
quenching. In this framework the leading nonanalytic dependence of the heavy 
baryon masses on the light quark masses is studied. The size of quenching 
effects is estimated by comparing the results of quenched and ordinary chiral 
perturbation theories. It is found that in general they can be large. This 
estimate is relevant to lattice simulations of the heavy baryon masses.    
\end{abstract}


\section{Introduction}

Recent technological developments have made it possible to perform lattice 
simulations of QCD with increased accuracy.  At the same time the precision of 
the lattice calculations depends on the understanding of the errors due to 
various approximations used in lattice simulations. One of the most important 
errors is due to quenching. The modifications induced by quenching at short 
distances have been studied in systems of quarkonia~\cite{El-Khadra:alphas1}
and are more or less well understood. 
Because of its nonperturbative nature, the long distance effects of quenching 
are more difficult to quantify. One  
approach is to use Quenched Chiral Perturbation Theory (QChPT). The idea
was first proposed by Sharpe~\cite{Sharpe:QChPT1,Sharpe:QChPT2}, with the 
formalism further developed by Bernard and 
Golterman~\cite{BernGolt:QChPT1,BernGolt:QChPT2}. The 
advantage of using the quenched version of Chiral Perturbation Theory (ChPT)
is that its predictions follow from the basic properties of the underling 
theory, and ChPT describes the low energy dynamics of ordinary QCD very well.

More recently, QChPT has been extended to describe the interactions of soft
pions with baryons~\cite{Labrenz:QBary1,Labrenz:QBary2}, heavy mesons~\cite{Booth:QChHQ,Sharpe:QHQ} and vector mesons~\cite{Booth:QChVE}. Here it is further 
extended to describe the interactions with heavy baryons. Ordinary 
chiral perturbation theory for heavy baryons has been formulated  by a
number of authors~\cite{Wise:HQET,Yan:HQChPT,Cho:HQ1}. We adapt their 
formalism to quenched QCD.

As every effective theory, ChPT contains a number of undetermined couplings all
of which must be fixed from experiment. Only for the self-interactions of
pions are there enough data to carry out this procedure beyond the leading 
order. The situation is even worse in the quenched case, where the only ``data''
available are extracted from lattice simulations. For this very reason we cannot
obtain any accurate quantitative predictions from our 
calculations. Nevertheless, there are certain loop corrections whose existence
is predicted unambiguously by the lowest order Lagrangian, in particular, terms 
that have nonanalytic dependence on the light quark masses. Contributions of 
these terms to the matrix elements cannot be absorbed into the higher order 
terms of the chiral expansion. Furthermore, since the nonanlytic dependence 
arises from the infrared region of the loops, it is particularly sensitive to 
the low energy behavior of the theory.

In this paper the nonanalytic corrections to heavy baryon masses are computed
in the framework of QChPT. Comparison with the analogous corrections obtained 
in ChPT shows the qualitative differences between the quenched and unquenched 
cases. It is also found that these nonanalytic terms could in general be large.
This is important for the lattice calculations of the heavy baryon masses, 
where results are extrapolated linearly in the light quark masses.

Since quenched QCD is not a unitary theory, it is possible that QChPT does not
describe its behavior at low energies correctly. However, the most
important qualitative feature of QChPT, the presence of quenched logarithms of 
the form $M_0^2 \ln{m_q^2}$, was first derived in strong-coupling perturbation 
theory ~\cite{Morel:QLogs,Sharpe:QChPT1}. Moreover, the coefficients of these 
terms obtained by the two different approaches are the same. We thus  expect 
that the description of the low energy limit of quenched QCD by QChPT is valid.

\section{Chiral Perturbation Theory for Heavy Baryons}

Heavy Quark Effective Theory (HQET) and ChPT are 
combined to describe the interactions of heavy baryons with soft Goldstone 
mesons. Here we give a brief review of the formalism.

To leading order of chiral expansion, self-interactions of soft 
pseudoscalar mesons are described by the Lagrangian

\begin{equation}
\label{eqn:CCWZ}
  \scr{L}= {f^2\over 8} \left(
   \Tr[\partial_\mu\Sigma\partial^\mu\Sigma^\dagger]
   +4B_0\,\Tr[M_{+}] \right)\, ,
\end{equation}
where $\Sigma = \xi^2$ and $\xi = e^{i\phi / f}$. Eight light pseudoscalar 
mesons are combined in a matrix  $\phi$,

\begin{equation}
   \phi = 
   \pmatrix{\frac 1{\sqrt 2}\pi^0 + \frac 1{\sqrt 6}\eta&
   \pi^+ & K^+ \cr
   \pi^- & -\frac 1{\sqrt 2} \pi^0 + \frac 1{\sqrt 6}\eta&
   K^0 \cr
   K^- & {\overline{K}}^0 & -\sqrt{\frac 2 3} \eta\cr} \,.
\end{equation}

The normalization is such that $f_\pi\approx130\MeV$. The leading symmetry 
breaking term depends on 

\begin{equation}
  M_{+} = \frac 12 (\xi^\dagger M\xi^\dagger + \xi M \xi)\, ,
\end{equation}
where $M$ is the quark mass matrix:

\begin{equation}
   M = 
   \pmatrix{m_u & 0 & 0 \cr
   0 & m_d & 0\cr
   0 & 0 & m_s \cr}\, . 
\end{equation}

The mass term of (\ref{eqn:CCWZ}) breaks the $SU(3)_L \times SU(3)_R$ symmetry 
which is restored in the limit $m_u ,m_d ,m_s \rightarrow 0$. In this limit the
eight pseudoscalar mesons become massless. Note that the meson masses are 
nonlinear in the quark masses.   

Heavy baryons contain one heavy quark, $c$ or $b$, and two light quarks.
According to HQET, in the leading order of heavy quark mass expansion, 
properties of the heavy baryons depend entirely on the configuration of light 
quarks and are independent of the spin and the flavor of the  heavy quark. The 
three
light quarks, $u\,, d$, and $s$, that form the fundamental representation of 
flavor $SU(3)$ can be combined in pairs to form an $SU_f (3)$ antitriplet and 
an $SU_f (3)$ sextet. There is a correlation between the spin wavefunction and 
the flavor wavefunction of the light quarks; those in  the $SU_f (3)$ 
antitriplet combine in a spin 0 state and those in  the $SU_f (3)$ sextet have 
spin 1. The 
former, combined with a spin 1/2 heavy quark, form a triplet of baryons with 
$J^P = {\frac{1}{2}}^+$. The spin 1
combination of light quarks together with the heavy quark forms two sextets of 
baryons with $J^P = {\frac{3}{2}}^+$ and $J^P = {\frac{1}{2}}^+$, which are
degenerate in the infinite heavy quark mass limit, $M_Q \rightarrow \infty$.
Since the configuration of light quarks in the  triplet baryons differs from that in the
sextet baryons, the two are not degenerate even in the $M_Q \rightarrow \infty$
limit.

The two sextets of baryons can be described with one field 
$S^{ij}_\mu$~\cite{Falk:HQH}. 

\begin{eqnarray}
S_\mu^{ij} &=& S^{ij}_{3/2 \,\mu} + S^{ij}_{1/2 \, \mu}  \nonumber \\
S^{ij}_{1/2 \, \mu} &=& \sqrt{\frac{1}{3}} \left( \gamma_\mu + v_\mu \right) 
\gamma^5 \frac{1+\vslash}{2} B_6^{ij} \\
S^{ij}_{3/2 \,\mu} &=& \frac{1+\vslash}{2}B^{\ast \, ij}_{6\mu} \nonumber \,, 
\end{eqnarray}
where $v^{\mu}$ is the 4-velocity of the baryon. The baryons with $J^P=\frac{1}{2}^{+}$ are combined in a matrix $B_6^{ij}$ 
\begin{equation}
   B_6^{ij} = 
   \pmatrix{\Sigma^{+1}_Q& \sqrt{\frac{1}{2}}\Sigma^0_Q 
            &\sqrt{\frac{1}{2}}{\Xi^{1/2}_Q}' \cr
   \sqrt{\frac{1}{2}}\Sigma^0_Q & \Sigma^{-1}_Q&
    \sqrt{\frac{1}{2}}{\Xi^{-1/2}_Q}'\cr
   \sqrt{\frac{1}{2}}{\Xi^{+1/2}_Q}'& \sqrt{\frac{1}{2}}{\Xi^{-1/2}_Q}' &
   \Omega_Q\cr}\, . 
\end{equation}
The matrix $B^{\ast}_{6\mu}$ is similar to $B_6$, except that it contains $J^P=
\frac{3}{2} ^{+}$ baryons. Here the notation of
Ref.~\cite{Yan:HQChPT} is used, where the superscripts denote the  $I_3$ 
projection of the isospin of the baryons. 

The baryons in the three dimensional representation of $SU_f (3)$ are described
by a field $T^{ij}$. 

\begin{equation}
T^{ij} = \frac{1+\vslash}{2} B_{\overline 3}^{ij}\,,   
\end{equation}

\begin{equation}
   B_{\overline 3}^{ij} = 
   \pmatrix{0& \Lambda_Q &\Xi^{+1/2}_Q \cr
   -\Lambda_Q & 0 & \Xi^{-1/2}_Q\cr
   -\Xi^{+1/2}_Q& -\Xi^{-1/2}_Q & 0 \cr} 
\end{equation}

To lowest order in the  $1/M_Q$ expansion, strong interactions of the heavy 
baryons with the soft pions are described by the chiral Lagrangian 
\footnote{The normalization and the notation for coupling constants are the same
as in~\cite{Savage:HBM}.} 

\begin{eqnarray}
  {\cal L}_{\rm kin} &=&  \frac{i}{2} \, \tr [\Tbar \left( v \cdot
       {\cal D} \right) T ] 
  -i \,  \tr [\Sbar ^\mu \left( v \cdot{\cal D} \right) S _\mu] \nonumber \\
  {\cal L}_{\rm mass} &=& \frac{\Delta _0}{2}\, \tr[\Tbar T] + 
  \lambda_1 \,\tr[\Sbar ^\mu  M_+ S _\mu] +  
  \lambda_2 \, \tr[\Sbar ^\mu S _\mu]\tr[M_+ ] \nonumber \\
   && \mbox{} + \frac{\lambda_3}{2} \, \tr[\Tbar M_+ T ] + 
  \frac{\lambda_4}{2} \, \tr[\Tbar T ] \tr[M_+ ] \nonumber \\
  {\cal L}_{\rm int} &=& 
  g_3\, \left( \tr[\Tbar A^\mu S_\mu] + {\rm h.c.} \right) 
  + ig_2 \tr[\Sbar ^\mu A^\rho S^\sigma ]\,
   v^\nu\epsilon_{\mu\nu\rho\sigma} \,. 
 \end{eqnarray}

Here 

\begin{equation}
\label{eqn:COV}
   D_\mu  S^{ij}_\nu = \partial_\mu S^{ij}_\nu + 
 {(V_\mu)}^i_k S^{kj}_\nu + {(V_\mu)}^j_k S^{ik}_\nu  
\end{equation}
is a covariant derivative, with a vector current

\begin{equation}
\label{eqn:VEC}
  V_\mu  = \frac 1 2
  \left(\xi \partial_\mu \xi^\dagger +\xi^\dagger \partial_\mu \xi \right)\,,
\end{equation}
and the axial current, $A_\mu$, is defined as follows:

\begin{equation}
\label{eqn:AXI}
  A_\mu =
  \frac i 2 \left(\xi \partial_\mu \xi^\dagger -\xi^\dagger \partial_\mu
  \xi \right)\,.
\end{equation} 
The covariant derivative of $T^{ij}$ has the same form.

We chose to absorb the mass of the sextet baryons  into the static phase of the
heavy baryon fields. The mass splitting between the sextet and triplet baryons,
finite as $M_Q \rightarrow \infty$, is denoted by $\Delta_0$. The propagators 
for $J^P ={\frac{3}{2}}^+$ and $J^P ={\frac{1}{2}}^+$ sextet baryons are 

\begin{equation}
i\, \frac{1+\vslash}{2} \, \frac{ - g^{\mu \nu} +v^\mu v^\nu + \frac{1}{3} 
(\gamma^\mu - v^\mu)(\gamma^\nu + v^\nu)}{v\cdot k} 
\end{equation}
and 

\begin{equation}
-i\,  \frac{1+\vslash}{2} \, \frac{\frac{1}{3}(\gamma^\mu - v^\mu)(\gamma^\nu 
+ v^\nu)}{v\cdot k}\, , 
\end{equation}
respectively, while the propagator of the triplet baryon is

\begin{equation}
\frac{1+\vslash}{2} \frac{i}{v\cdot k + \Delta_0}\,. 
\end{equation}

\section{Quenched QCD}
   
Quenched QCD is obtained from the ordinary theory by removing the 
disconnected fermion loops. Formally this is done by introducing a  bosonic 
``ghost'' partner  $\tilde q^i$~\cite{Morel:QLogs} for each quark $q^i$, which 
has the same mass and couples to the gluons the same way that the quark does. 
The sign difference between the fermionic and the bososnic loops results in 
cancellation of the two.  The symmetry of the quenched theory is enlarged from 
$SU(3)_L \times SU(3)_R$ to the semi-direct product  
$(SU(3|3)_L \times SU(3|3)_R){\otimes}U(1)$. 

In this larger theory the Goldstone matrix becomes a supermatrix,

\begin{equation}
   \Pi = \left(\matrix{\pi&\chi^\dagger\cr
          \chi&\tilde\pi\cr}\right),
\end{equation}
where the quark/ghost content of the fields is
$\pi\sim q\qbar$, $\chi^\dagger\sim\qt\qbar$, $\chi\sim q\qtbar$ and
$\tilde\pi\sim\qt\qtbar$.  Each of these is an $3\times 3$ matrix; for example 
\begin{equation}
   \pi = 
   \pmatrix{\frac 1{\sqrt 2}\pi^0 + \frac 1{\sqrt 6}\eta&
   \pi^+ & K^+ \cr
   \pi^- & -\frac 1{\sqrt 2} \pi^0 + \frac 1{\sqrt 6}\eta&
   K^0 \cr
   K^- & {\overline{K}}^0 & -\sqrt{\frac 2 3} \eta\cr} 
   + \frac1{\sqrt 3}\, \eta'I_3\,.
\end{equation}
Note that $\chi$ and $\chi^\dagger$ are fermionic fields, while $\pi$
and $\tilde\pi$ are bosonic. 

Notice one important difference from ordinary ChPT: the inclusion of the 
$SU(3)$ singlet in the theory. In ordinary QCD the anomaly pushes the 
mass of the  $\eta'$ up to the chiral scale and it decouples from the low 
energy effective theory. In the quenched case, because of the absence of 
disconnected quark loops, this decoupling does not occur and the $\eta'$ has to be included in the theory. In the quenched case additional terms are required 
to describe the dynamics of the anomalous field. The lowest order Lagrangian is
similar to Eq.~(\ref{eqn:CCWZ}):

\begin{eqnarray}\label{eqn:LBG}
   \scr{L}_{Q\chi} &=& {f^2\over 8} \left(
   \Str[\partial_\mu\Sigma\partial^\mu\Sigma^\dagger]
   +4B_0\,\Str[\scr{M}_{+}] \right)\nonumber\\
   &&\qquad+{1\over2}\left(A_0\Str[\partial_\mu\Pi]
   \Str[\partial^\mu\Pi]- M_0^2 \Str[\Pi]\Str[\Pi]\right)\,,
\end{eqnarray}
with $\Sigma = \xi^2$, $\xi = e^{i\Pi / f}$ and 
\begin{equation}
  \scr{M} = \left(\matrix{M&0\cr 0&M\cr}\right)\,.
\end{equation}

$\scr{M}_{+}$ is defined analogously to $M_{+}$. The ``supertrace'' $\Str$ is 
defined with a minus sign for the ghost-antighost fields. Normalization of 
$A_0$ and $M_0$ is such that they have no implicit dependence on $N_f$, the 
number of flavors.

The presence of the $\Str[\Pi]=N_f^{1/2}\,(\eta'-\tilde\eta')$ term leads to a
double-pole structure for the propagators of the flavor-neutral mesons. In
the basis where these mesons correspond to $q_i\qbar_i$ and $\qt_i\qtbar_i$ this propagator takes the form 
\begin{equation}
  G_{ij}(p) = {\delta_{ij} \epsilon_i \over p^2 - \mii^2} 
  + {-A_0 p^2 + M_0^2 \over (p^2 - \mii^2)(p^2 - \mjj^2)}\,,
\end{equation}
where $\mii^2 = 2B_0 m_i$, and $\epsilon_i=1$ if $i$ corresponds to a
quark and $\epsilon_i=-1$ if $i$ corresponds to a ghost. The second term in the
propagator can be treated as a vertex, called a hairpin, with the rule that it 
can be inserted only once on a given meson line.

Inclusion of quenching effects in the heavy baryon sector is straightforward. 
The baryon matrices are promoted to supermatrices and $\Tr$ is replaced by 
$\Str$. The kinetic and the mass terms of the quenched Lagrangian of the heavy
baryons are

\begin{eqnarray}
\label{eq:QLHB}
  {\cal L}^{Q\chi}_{\rm kin} &=&  \frac{i}{2} \, \Str [{\cal \Tbar} 
  \left( v \cdot {\cal D} \right) {\cal T} ] 
  -i  \Str [{\cal \Sbar} ^\mu \left( v \cdot{\cal D} \right) 
      {\cal S} _\mu] \, ,\nonumber  \\
  {\cal L}^{Q\chi}_{\rm mass} &=& \frac{\Delta _0}{2}\, \Str[{\cal \Tbar T}] + 
  \lambda_1 \,\Str[{\cal \Sbar} ^\mu {\cal M}_+ {\cal S} _\mu] +  
  \lambda_2 \, \Str[{\cal \Sbar} ^\mu {\cal S} _\mu ]\Str[{\cal M}_+ ] 
    \nonumber  \\
  &&\mbox{} + \frac{\lambda_3}{2} \, \Str[{\cal \Tbar} {\cal M}_+ {\cal T} ] + 
  \frac{\lambda_4}{2} \, \Str[{\cal \Tbar T} ] \Str[{\cal M}_+ ]\,.
 \end{eqnarray} 
where
\begin{equation}
{\cal S} _\mu =  \left(\matrix{S_\mu & K_\mu \cr
          L_\mu &{\tilde S}_\mu \cr}\right)\, ,  
\end{equation}
with $K_\mu$ and $L_\mu$ describing the heavy baryons obtained by replacing one
of the quarks by its ``ghost'' partner and ${\tilde S}_\mu$ is the heavy 
baryon with both light quarks replaced by ``ghosts''.

The superfield ${\cal T}$ is defined the same way,
\begin{equation}
{\cal T}  =  \left(\matrix{T & R\cr
          P &\tilde T\cr}\right) \,.  
\end{equation}

The covariant derivative, the vector and the axial currents are defined 
analogously to ~(\ref{eqn:COV})  --~(\ref{eqn:AXI}).

In the quenched case, $\Str {\cal A}_\mu$ does not vanish and the symmetry 
allows the baryon fields to couple to the flavor singlet Goldstone mesons 
through this term. Thus the interaction part of the quenched Lagrangian has one
additional coupling that is absent in the unquenched case:

\begin{eqnarray}
{\cal L}^{Q\chi}_{\rm int} &=& 
  g_3\, \left( \Str[{\cal \Tbar} {\cal A}^\mu {\cal S}_\mu] + {\rm h.c.} \right) \nonumber\\
  &&\mbox{}+
  ig_2 \Str[{\cal \Sbar} ^\mu {\cal A}^\rho {\cal S}^\sigma ]\,
   v^\nu\epsilon_{\mu\nu\rho\sigma} \nonumber\\
  && \mbox{}+
  ig_1 \Str[{\cal \Sbar} ^\mu {\cal S}^\sigma]\Str[{\cal A}^\rho]\, 
        v^\nu\epsilon_{\mu\nu\lambda\sigma}\,.
\end{eqnarray}

\section{Nonanalytic corrections to heavy baryon masses}

We encounter two types of integrals in our calculations. One, $J^{\mu \nu}_1$, 
arises from the chiral loops without the  hairpin vertex and the other,  
$J^{\mu \nu}_2$, from the loops with the hairpin vertex on Goldstone meson 
line:
\begin{equation}
\label{eq:INT1}
J^{\mu \nu}_1 (m,\Delta_0, v \cdot k) = i\int \frac{d^n p}{{(2\pi )}^n f^2}
\frac{p^\mu p^\nu}{(p^2 - m^2 + i\epsilon )(v \cdot (p+k) + \Delta_0 + 
i\epsilon )}   
\end{equation}
and 
\begin{equation}
\label{eq:INT2}
J^{\mu \nu}_2 (x,y,\Delta_0, v \cdot k) = \frac{1}{x^2 - y^2} \left[ 
(M_0^2 - A_0 x^2 )J^{\mu \nu}_1 (x,\Delta_0, v \cdot k) -    
(M_0^2 - A_0 y^2 )J^{\mu \nu}_1 (y,\Delta_0, v \cdot k) \right]\,.
\end{equation}

We truncate the series to include terms of order  
$m_q$, $m_q \log m_q$ and $m_q^{3/2}$ but not those of order $m_q^2$ and 
higher. If not otherwise indicated, dimensionful parameters $\Delta _0$ and
$M_0^2$ are considered to be of order $m_q$. With these assumptions, terms 
contributing to the baryon mass corrections are:\footnote{Setting $\Delta_0 =0$
in  (\ref{eq:INT1}) and (\ref{eq:INT2}) means that we neglect terms of order
$\Delta_0 m_q \ln{m_q}$.}

\begin{eqnarray}
  J^{\mu \nu}_1 (m, 0, 0) &=& (v^\mu v^\nu -g^{\mu \nu })I_1 (m)  \\
  J^{\mu \nu}_2 (x, y, 0,0) &=& (v^\mu v^\nu -g^{\mu \nu })I_2 (x,y)
  \nonumber \,,
\end{eqnarray}
where
\begin{eqnarray}
 \label{eq:1int} 
  I_1 (x) &=& -\frac{x^3}{12\pi f^2}  \\
  I_2 (x,y) &=& \frac{1}{x^2 - y^2}\left[ (M_0^2 - A_0 x^2 )I_1 (x) - 
  (M_0^2 - A_0 y^2 )I_1 (y) \right] \nonumber \,.
\end{eqnarray}
$I_2 (x,y)$ has the limit
\begin{equation}
  \label{eq:Limit}
  I_2 (x,x)=-\frac{x^3}{12\pi f^2} \left( \frac{3}{2} M_0^2 x 
-\frac{5}{2}A_0 x^3 \right)\,.
\end{equation}

\begin{figure}
\epsfxsize=12cm
\hfil\epsfbox{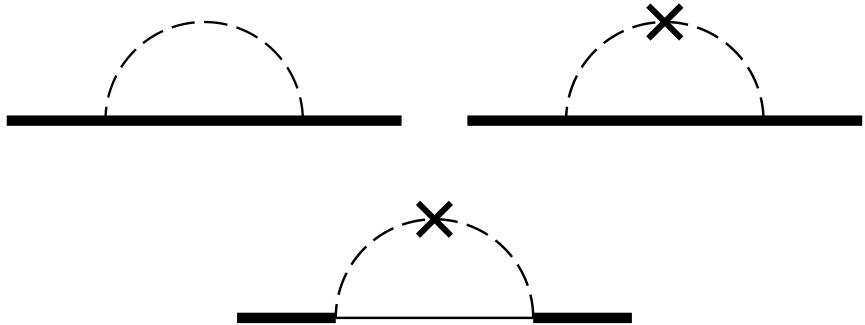}\hfill
\caption{Diagrams that modify the masses of sextet baryons with light quarks of
two different flavor. The thin solid line denotes the triplet baryons.}
\label{fig:2sixtet}
\end{figure}

Here we only compute corrections to masses arising from the chiral loops and 
not from the $1/M_Q$ expansion. Chiral corrections do not lift the degeneracy 
between the $B_6$ and $B_6^*$ baryons. However, the baryon masses inside the 
sextets as well as inside the triplet are split. Mass corrections to the 
baryons that have the same flavor light quarks differ from those that have  
light quarks of two different flavors.
The latter get contributions from loops which contain triplet baryons. This
is different from the unquenched case, where the loops with the triplet baryons
contribute in both cases. The reason for this difference is that the diagrams 
with the triplet baryons which do not contain the hairpin vertex require a 
closed quark loop, and in the quenched theory they vanish. On the other hand, 
the hairpin vertex appears only on the flavor diagonal meson line and these 
mesons only couple the triplet baryons with the off-diagonal sextet baryons.  
The diagrams that modify sextet baryon masses are shown 
in Fig.~\ref{fig:1sixtet} and Fig.~\ref{fig:2sixtet}. The contributions from 
these diagrams are:

\begin{figure}
\epsfxsize=12cm
\hfil\epsfbox{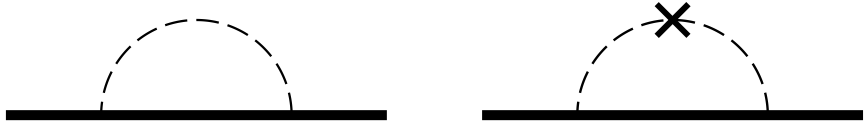}\hfill
\caption{Diagrams contributing to the flavor diagonal sextet baryon mass 
corrections. The thick solid line denotes the sextet baryons and  the dashed 
line denotes the Goldstone mesons.}
\label{fig:1sixtet}
\end{figure}

\begin{equation}
 \label{eq:mii} 
  \delta M^6_{ii} =  2g_2 g_1 I_1(\mii) + g_2^2 I_2(\mii , \mii) \, ,
\end{equation}
for the diagonal baryons, and
\begin{eqnarray}
  \label{eq:mij}
  \delta M^6_{ij} &=&  g_2 g_1 \left( I_1(\mii) + I_1(\mjj) \right) +
  \frac{1}{4} g_2^2 \left( I_2(\mii , \mii) +  I_2(\mjj , \mjj) + 
    2 I_2(\mii , \mjj) \right) \nonumber \\
                &&+ \frac{1}{4} g_3^2 \left( I_2(\mii , \mii) +  
    I_2(\mjj , \mjj) - 2 I_2(\mii , \mjj) \right) \,
\end{eqnarray}
for the off-diagonal sextet baryons. $I_1 (x)$ and $I_2 (x,y)$ are defined in 
Eq.(\ref{eq:1int}). In the limit of exact isospin symmetry the term 
proportional to $g_3^2$ in Eq.(\ref{eq:mij}) vanishes when the baryon 
contains  up and  down quarks. 

\begin{figure}
\epsfxsize=6cm
\hfil\epsfbox{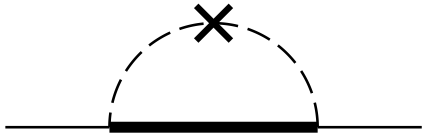}\hfill
\caption{Diagram that modifies the masses of triplet baryons.}
\label{fig:triplet}
\end{figure}

As was discussed above, in the case of triplet baryons the only diagrams that 
contribute to the mass corrections are those with the hairpin vertex 
(Fig.~\ref{fig:triplet}). The triplet baryon mass corrections have the 
following form:   

\begin{equation}
  \delta M^3_{ij} =  \frac{3}{4}  g_3^2 \left( I_2(\mii , \mii) +  
I_2(\mjj , \mjj) - 2 I_2(\mii , \mjj) \right)
\end{equation}

This expression has the same form as the correction to the off-diagonal sextet
baryon masses that arise from the diagrams with the triplet baryons. Thus in
the limit of exact isospin symmetry the mass corrections to  
$\Lambda_Q$ baryon vanish.

There is no mixing between the triplet and the sextet baryons in the leading 
order of the  $1/M_Q$ expansion. Mixing would require the breaking of heavy
baryon spin symmetry, which is exact at this order of the heavy quark 
expansion.

Finally, we note that these results are readily generalized to an arbitrary 
number of light flavors.

\section{Phenomenology}

As was mentioned in the introduction, there are many undetermined 
parameters in the theory. In the quenched case there are additional parameters 
that are absent from the Lagrangian of  ordinary ChPT. Of course the
parameters of the  quenched and unquenched theories, even those that are 
present in both cases, might not be the same. Here couplings that are 
present in ordinary ChPT are assumed to have the same value in the quenched case. Other 
couplings we assume to be of order one.   Calculations are done for two sets of
values for $M_0$ and $A_0$: $M_0 =400$ \mev, $A_0 =0$ and $M_0 =100$ \mev, 
$A_0 =0.2$.  For more detailed discussion of the values of these 
parameters we refer the reader to~\cite{Booth:QChVE}.       

There is a nontrivial relation between the sextet baryon masses that holds in
both quenched and unquenched cases in the leading order of $SU_f (3)$ breaking.
Namely, in the limit of exact isospin symmetry:

\begin{equation}
\delta M=M_{\Sigma_Q} + M_{\Omega_Q} - 2M_{\Xi_Q'} = 0\,.
\end{equation}
 
This relation gets corrections in the next order of the chiral expansion. In 
the limit of vanishing $\Delta_0$, this correction is~\cite{Savage:HBM}

\begin{equation}
M_{\Sigma_Q} + M_{\Omega_Q} - 2M_{\Xi_Q'} = \frac{1}{24\pi f^2}(g_2^2 - g_3^2)
(4m_K^3 - 3m_{\eta}^3 - m_{\pi}^3)\,,
\end{equation}
assuming the tree level relation between the Goldstone mesons, $4m_K^2 -
3m_{\eta}^2 - m_{\pi}^2 =0$. In the quenched case the equivalent relation 
takes the form 

\begin{equation}
M_{\Sigma_Q} + M_{\Omega_Q} - 2M_{\Xi_Q'} = \frac{1}{2} (g_2^2 - g_3^2)
\left[ I_2(\muu , \muu) + I_2(\mss , \mss) -2I_2(\muu , \mss) \right]\,.
\label{eq:DLT}
\end{equation}

To simplify the last relation let us consider the limit $m_u =m_d =0$. Then
Eq.~(\ref{eq:DLT}) becomes:
\begin{equation}
M_{\Sigma_Q} + M_{\Omega_Q} - 2M_{\Xi_Q'} = \frac{1}{48\pi f^2} (g_2^2 - g_3^2)
\left[ M_0^2 \mss + A_0 \mss^3 \right].
\end{equation}
  
\begin{figure}
\epsfxsize=12cm
\hfil\epsfbox{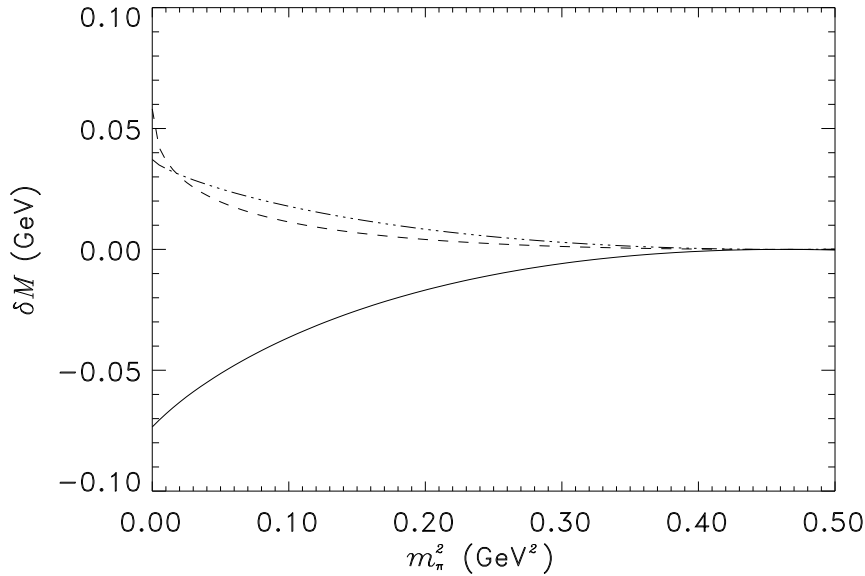}\hfill
\caption{Dependence of $\delta M$ on $m_{\pi}^2$ in the quenched and unquenched
cases. The solid line corresponds to the unquenched case, the dashed line to 
the quenched case with $M_0=400\mev$ and $A_0=0$, and the dashed-dotted line to
the quenched case with $M_0=100\mev$ and $A_0=0.2$.}
\label{fig:NTR}
\end{figure}

The main difference between the two cases is the presence of the term 
proportional to $m_q^{1/2}$ in the quenched case. The Goldstone meson masses 
are linear in $m_q^{1/2}$. Fig.~\ref{fig:NTR} shows the dependence of
these corrections on $m_{\pi}^2$ while keeping the value of the strange
quark mass fixed. (This is the extrapolation used in lattice calculations to 
extract the values of the baryon masses.) In both cases the corrections to 
$\delta M$ are small but they have different signs. The plot indicates the 
very different behavior of the nonanalytic corrections in the two cases.  

We also investigate the mass splittings between the sextet and the triplet 
baryons. The splittings are calculated in the limit of exact isospin 
symmetry.\footnote{In this limit $\muu =\mdd =m_{\pi}$.} As was discussed in 
the previous section, in this limit $\Lambda_Q$ 
does not receive corrections at the first order of the chiral expansion. 
The results for the mass splittings are:\footnote{Not included in the following
expressions are the contributions from the mass terms of the Lagrangian 
(\ref{eq:QLHB}) that are linear in the light quark masses. These terms depend 
on the undetermined parameters, $\lambda_i$. We found that these contributions
are small if one considers the linear combinations of $\lambda_i$'s to be of 
order 1.} 

\begin{equation}
 \label{eq:Split1}
 \Sigma_Q -  \Lambda_Q = \Delta_0 - 2g_2 g_1 \frac{m_{\pi}^3}{12\pi f^2}  - 
 g_2^2 \frac{1}{12\pi f^2}\left( \frac{3}{2}M_0^2 m_{\pi} - \frac{5}{2}
 A_0 m_{\pi}^3 \right)   
\end{equation}
and
\begin{eqnarray}
\Xi'_Q -  \Xi_Q &=& \Delta_0 + g_2 g_1 (I_1(m_{\pi}) + I_1(\mss))+ \frac{1}{4}
g_2^2(I_2(m_{\pi} , m_{\pi}) + I_2(\mss ,\mss ) + 2I_2(m_{\pi} ,\mss)) 
\nonumber \\
&&- \frac{1}{2} g_2^2 (I_2(m_{\pi} , m_{\pi}) + I_2(\mss ,\mss ) - 
2I_2(m_{\pi} ,\mss))\,.
\end{eqnarray}
The last relation simplifies in the $m_u =m_d =0$ limit: 
\begin{equation}
\label{eq:Split2}
\Xi'_Q -  \Xi_Q =\Delta_0 - \frac{M_0^2 \mss}{48\pi f^2} (\frac{7}{2} g_2^2 +
g_3^2) - g_2 g_1\frac{\mss^3}{12\pi f^2}\,.
\end{equation}
For simplicity of the analytic expressions only the cases with $A_0 =0$ are
shown.

\begin{figure}
\epsfxsize=12cm
\hfil\epsfbox{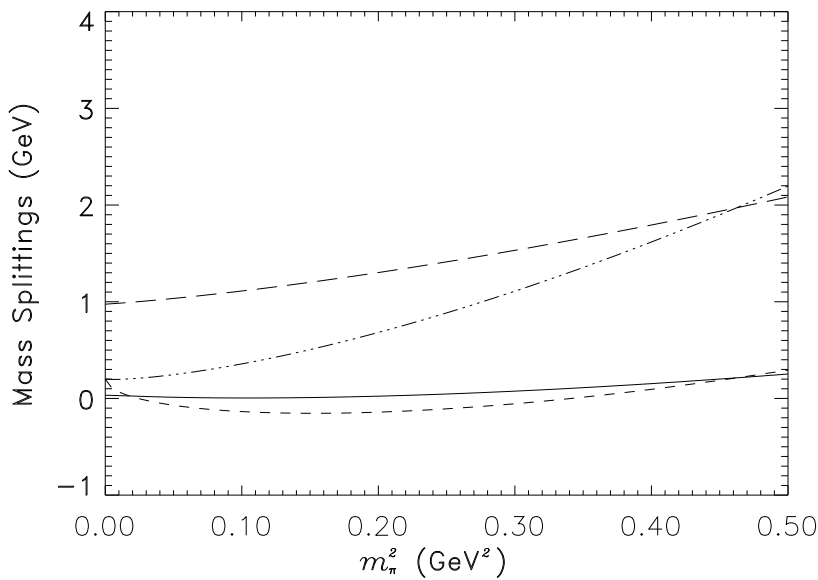}\hfill
\caption{Dependence of the sextet--triplet mass splittings on $m_{\pi}^2$. The 
solid and dashed lines correspond to the $\Xi'_Q -  \Xi_Q$ splitting with 
$M_0 =400\mev \, , A_0=0$ and with $M_0 =100\mev \, , A_0=0.2$ respectively. 
The line with small dashing and the dashed-dotted line correspond to the 
$\Sigma_Q -  \Lambda_Q$ splitting with $M_0 =400\mev \, , A_0=0$ and with 
$M_0 =100\mev \, , A_0=0.2$ respectively. $g_1$ is taken to be 0.5.}
\label{fig:SPT}
\end{figure}

We would like to point out two things. First, the corrections to the mass 
splittings, as the correction to $\delta M$, display  nonanalytic 
dependence on the light quark masses that is different from ordinary ChPT. 
In the quenched case the leading nonanalytic term is $m_q^{1/2}$. Notice that 
the coefficient of the term linear in the Goldstone meson mass depends, apart 
from the $M_0^2$ term, on the parameters that exist in the unquenched theory. 
One can estimate the size of this term assuming that the corresponding 
parameters in the quenched and unquenched theories are the same. Second, there 
are terms proportional to $g_1$, the coupling that is absent from the ordinary 
theory. In general this term gives a large contribution to the mass 
corrections,
which depends on the value of $g_1$, of course. This makes the 
estimate of quenching errors even more uncertain. The dependence of the mass splittings on  $m_{\pi}^2$ for the
fixed strange quark mass is shown in Fig.~\ref{fig:SPT}. Here $g_1$ is 
taken to be 0.5 and $\Delta_0 =200\mev$. 
Note that, by contrast, the terms proportional 
to $g_1$ are absent from the corrections to $\delta M$. In this particular 
combination of the heavy baryon masses these terms cancel. This indicates that
one has to be careful when using such mass relations for the estimate of 
quenching errors.  
  
Recently these splittings, as well as the heavy baryon masses, were calculated 
on the lattice~\cite{UKQCD:Lat1}. In that analysis, calculations are performed 
for the light quark masses of order of
the strange quark mass, and then the results are linearly extrapolated to 
the chiral limit of the up and down quark masses, while the strange quark mass 
is fixed to its phenomenological value. In particular, the following relation
for the mass of the heavy baryon is assumed:

\begin{equation}
  \label{eq:lamass}
  M_{Qij}=\mu (M_Q) + C(m_i + m_j )\,,
\end{equation}
where $M_Q$ is the mass of the heavy  quark and $m_i$ and $m_j$ denote the 
masses of the light quarks. $\mu (M_Q)$ is the mass of the baryon in the chiral
limit, $m_u =m_d =m_s =0$. $\mu$ and $C$ are different for the sextet and 
triplet baryons. The mass splitting between the sextet and triplet baryons is:

\begin{equation}
  \label{eq:lsplit1}
  \Delta M_{Qij}=\Delta_0 (M_Q) + D(m_i + m_j )\,.
\end{equation}

From Eq.~(\ref{eq:lsplit1}) it follows that in the chiral limit for up and down
quarks $m_u =m_d =0$,

\begin{equation}
  M_{\Sigma_Q} - M_{\Lambda_Q}=\Delta_0 \,,
\end{equation}
and this agrees with Eq.~(\ref{eq:Split1}) when $m_{\pi} =0$. In ChPT 
$\Delta_0$ is a free parameter and has to be determined from the fit to the 
experimental data. For our choice of $\Delta_0 = 200\mev$ the quenched result 
is in agreement with the lattice result.  

The situation is different for the $\Xi'_Q - \Xi_Q$ mass splitting. The 
quenched result is given by Eq.~(\ref{eq:Split2}). The lattice result is

\begin{equation}
  \label{eq:Lsplit2}
  M_{\Xi'_Q} - M_{\Xi_Q} =\Delta_0 (M_Q) + Dm_s
\end{equation}
In this case the finite value of the strange quark mass results in the
modification of the tree level splitting even in the limit $m_u =m_d =0$. In 
QChPT the $\Xi'_Q - \Xi_Q$ splitting, in contrast to the 
$\Sigma_Q -  \Lambda_Q$ splitting, has different values for the two sets of 
$M_0$ and $A_0$ parameters in this limit. As Fig.~\ref{fig:SPT} indicates, the 
corrections to the $\Xi'_Q - \Xi_Q$ splitting could in general be large. The
plot also shows that the size of the corrections strongly depends on the 
parameters $M_0$ and $A_0$.

The most important result, however, is the different dependence of the lattice 
and quenched results on the light quark masses. The nonanalytic terms, present 
in QChPT, are not accounted for in lattice simulations. The presence of these
nonanlytic terms is unambiguously predicted by the quenched theory and they can
be calculated from the tree level Lagrangian. Because of the presence of 
several undetermined parameters, it  is possible that the coefficients of the 
nonanalytic terms are small. The only way to determine the size of these terms
is to take them into account in lattice simulations and perform the nonlinear
extrapolations.

\section{Summary}

Let us  summarize briefly the results of our calculation. The main qualitative 
result is the dependence of the heavy baryon masses on {\it the square root} of
the light quark mass $m_q^{1/2}$. This is different from  ordinary ChPT 
where the leading nonanalytic dependence goes as $m_q^{3/2}$. The presence of 
these terms is unambiguously predicted. Their coefficients are calculated from
the tree level Lagrangian and they can not be modified by the inclusion of the 
higher order terms in the chiral expansion. Because the coefficients in the 
Lagrangian for QChPT are not well constrained, it is difficult to make 
numerical predictions of the size of the nonanalytic terms. However, it is 
certain that lattice simulations should be modified to account for this leading
nonanalytic behavior.

\section{Acknowledgments}
  
I am greatly indebted to Adam Falk, Michael Booth and Thomas Mehen 
for many useful discussions on the subject. This work was supported by the 
National Science Foundation under Grant No. PHY--9404057. 

\bibliography{tramali}
\bibliographystyle{prsty}

\end{document}